\documentclass[reprint,superscriptaddress,aps,prd,floatfix]{revtex4-1}
\usepackage[a4paper,left=1.5cm,right=1.5cm,top=3cm,bottom=3cm]{geometry}
\usepackage[toc]{appendix} 

\usepackage{graphicx}
\usepackage{dcolumn}
\usepackage{bm}
\usepackage{xcolor}
\usepackage{amssymb,amsmath,amsfonts,physics}
\usepackage{graphicx}
\usepackage{longtable}
\usepackage{verbatim}
\usepackage{color}
\usepackage{orcidlink}  
\usepackage{mdframed}
\usepackage{placeins}
\usepackage{soul}
\usepackage{amsfonts,amssymb,mathrsfs,amsmath,esint}
\usepackage{slashed, cancel}
\usepackage{framed}
\usepackage{mdframed}
\usepackage{simplewick} 
\allowdisplaybreaks
\usepackage{latexsym}
\usepackage{graphicx}
\graphicspath{{./Figures/}}
\usepackage{booktabs}
\usepackage{datetime}
\usepackage{multirow}
\newdateformat{mydate}{\THEDAY{ }\monthname[\THEMONTH]{ }\THEYEAR}

\usepackage{tikz}
\usepackage{color}
\usepackage{framed}
\usepackage{hyperref}
\usepackage[capitalise]{cleveref}
\hypersetup{colorlinks, citecolor=bluscuro, linkcolor=black, urlcolor=bluscuro}
\definecolor{rossos}{cmyk}{0,1,1,0.55}
\definecolor{bluscuro}{rgb}{0.15, 0.2, .85}
\definecolor{bluchiaro}{cmyk}{1,.3,0.,0.1}

\graphicspath{{./images/}}

\newcommand{\be}{\begin{equation}}
	\newcommand{\ee}{\end{equation}}
\newcommand{\bea}{\begin{eqnarray}}
	\newcommand{\eea}{\end{eqnarray}}
\newcommand{\beq}{\begin{equation}}
	\newcommand{\eeq}{\end{equation}}

\def\beqa{\begin{eqnarray}}

	\def\eeqa{\end{eqnarray}}

\def\lsim{\mathrel{\rlap{\lower4pt\hbox{\hskip0.5pt$\sim$}}
		\raise1pt\hbox{$<$}}}         
\def\gsim{\mathrel{\rlap{\lower4pt\hbox{\hskip0.5pt$\sim$}}
		\raise1pt\hbox{$>$}}}         

\newcommand{\bk}{\mathbf{k}}

\usepackage[normalem]{ulem}
\usepackage{soul}

\begin{document}
	
	\title{Primordial Gravitational Wave Background as a Probe of the Primordial Black Holes}
	
	\author{Utkarsh Kumar \orcidlink{0000-0002-8158-2078}}
	\email{utkarshkumar.physics@gmail.com}
	\affiliation{Department of Physics, University of Ottawa, Ottawa, ON K1N6N5, Canada}
	\date{\today}

\begin{abstract}
We study the formation of primordial black holes (PBHs) from the collapse of density perturbations induced by primordial gravitational waves (PGWs). The PGWs' interpretation of the stochastic gravitational wave background (SGWB) detected by the Pulsar Timing Array (PTA) corresponds to PBHs formation in the mass range $[10^{-12}-10^{-3}] M_{\odot}$. Importantly, our analysis shows that PGWs' interpretation of recent PTA data remains viable, as it does not lead to PBH overproduction. We derive the amplitude of PGWs by leveraging existing constraints on the PBH abundance across a wide mass range. Notably, these constrained amplitudes predict SGWB signals that would be detectable by future gravitational wave observatories.
\end{abstract}

\maketitle
	
\textbf{Introduction}-- The various detections of gravitational waves (GWs) from binary black holes (BBHs) \cite{LIGOScientific:2016aoc,LIGOScientific:2016sjg,LIGOScientific:2017bnn,LIGOScientific:2017vox,LIGOScientific:2017ycc} and a binary neutron star (BNS) \cite{LIGOScientific:2017vwq} by LIGO/Virgo have prompted interest in the physics of Primordial Black Holes (PBHs) \cite{Zeldovich:1967lct,Hawking:1971ei,Carr:1974nx,Carr:1975qj}. The PBH scenario is intriguing because it can explain the event rate of LIGO/Virgo BBHs and account for all dark matter \cite{GarciaBellido:1996qt,Khlopov:2008qy,Frampton:2010sw,Kawasaki:2016pql,Cotner:2016cvr,Cotner:2017tir,Carr:2016drx,Inomata:2016rbd,Pi:2017gih,Inomata:2017okj,Garcia-Bellido:2017aan,Inoue:2017csr,Georg:2017mqk,Inomata:2017vxo,Kocsis:2017yty,Ando:2017veq,Cotner:2016cvr,Cotner:2017tir,Cotner:2018vug,Sasaki:2018dmp,Carr:2018rid,Cotner:2019ykd}. There are several mechanisms for the formation of these PBHs, and their masses span a wide range of values \cite{Carr:2020xqk,Green:2020jor,Bird:2022wvk}. In the standard scenario, PBHs can form due to enhanced primordial curvature perturbations at small scales created by inflation in the very early Universe. These same curvature fluctuations can source tensor modes at second order in perturbation theory\cite{Baumann:2007zm,Picard:2023sbz,Domenech:2021ztg,Bari:2023rcw,Balaji:2023ehk,Domenech:2023jve,Chen:2024fir,Domenech:2024rks,Domenech:2024drm,Lozanov:2023aez,Lozanov:2023knf,Lozanov:2023rcd,Adshead:2021hnm,Domenech:2020kqm,Inomata:2019zqy,You:2023rmn,HosseiniMansoori:2023mqh,Gorji:2023sil,Liu:2023pau,Frosina:2023nxu,Choudhury:2023hfm,Harigaya:2023pmw,Kumar:2024bvp,Yogesh:2025,Gangopadhyay:2023qjr,Wang:2024vfv,Kim:2025dyi,Liu:2023ymk,Liu:2023hpw,Barroso:2024cgg,Papanikolaou:2020qtd,Papanikolaou:2022chm}. Consequently, PBH formation in this case inevitably leads to scalar-induced gravitational waves (SIGWs). Recent pulsar timing array (PTA) experiments, including NANOGrav \cite{NANOGrav:2023gor,NANOGRAV:2018hou}, CPTA \cite{Xu:2023wog}, PPTA \cite{Reardon:2023gzh}, and EPTA+InPTA \cite{EPTA:2023fyk,EPTA:2023xxk}, have shown evidence for the existence of a stochastic background of GWs (SGWB). If this observed SGWB is interpreted as SIGWs, it allows us to estimate the abundance of PBHs. Other possible sources of the observed SGWB, including phase transitions \cite{Ghosh:2023aum,Bringmann:2023opz,NANOGrav:2021flc,Gouttenoire:2023bqy,Chen:2023bms,Pritchard:2024vix,Salvio:2023blb,Salvio:2023ynn,Salvio:2023ynn,Salvio:2023blb,Salvio:2023ynn} that cause bubble collisions, cosmic string formation\cite{Ellis:2020ena,Blasi:2020mfx,Ellis:2023oxs,Maji:2023fhv}, and domain wall formation\cite{Gouttenoire:2023ftk,Zhang:2023nrs}, can produce populations of PBHs, except for primordial GWs (PGWs)\cite{Ben-Dayan:2023lwd,Vagnozzi:2023lwo,Benetti:2021uea,Vagnozzi:2022qmc,Das:2023nmm,Datta:2023vbs,Vagnozzi:2023lwo,Vagnozzi:2020gtf,Antoniadis:2022pcn,EPTA:2021crs,Goncharov:2021oub,NANOGrav:2020bcs,NANOGrav:2020bcs,Hobbs:2017oam,Renzini:2022alw,Ben-Dayan:2024aec,Ben-Dayan:2019gll,Kuroyanagi:2014nba,BICEP:2021xfz,Giare:2022wxq}.

In this paper, we explore an alternative approach to probe the PBH scenario in which PGWs act as a source for generating second-order radiation-density perturbations. This mechanism has been considered previously to study the generation and evolution of matter-density perturbations, quantifying the impact of PGWs on the Large-Scale Structure of the Universe \cite{Tomita1971,Tomita1972, Matarrese:1997ay,Bari:2021xvf,Bari:2022grh,Abdelaziz:2025qpn} as well as PBHs \cite{Nakama:2015nea,Nakama:2016enz}. Here, we focus on smaller scales relevant to current PTA data releases and next-generation gravitational wave interferometers, such as LISA \cite{LISACosmologyWorkingGroup:2022jok}, DECIGO \cite{Kawamura:2020pcg}, $\mu$Ares \cite{Sesana:2019vho}, and future LISA-type missions \cite{Ruan:2018tsw,Barke:2014lsa}.

The goal of this paper is to demonstrate that small-scale PGWs can produce PBHs across various mass windows. We focus specifically on inflation models where the linear power spectrum is a either power-law or has a log-normal shape, as predicted by several inflationary scenarios\cite{Guzzetti:2016mkm,Dimastrogiovanni:2016fuu,Thorne:2017jft,Namba:2015gja,Afzal:2024hwj,Maleknejad:2018nxz,Maleknejad:2016qjz}. We show that the PGW interpretation of NANOGrav and EPTA DR2 results is associated with PBH formation, with the mass function falling within $ \left[10^{-12} - 10^{-3} \right] M_{\odot} $. Additionally, our approach provides a framework to constrain tensor-mode amplitude by leveraging existing limits on PBH abundances set by evaporation \cite{Carr:2009jm}, gravitational lensing\cite{Niikura:2019kqi,Niikura:2017zjd,Barnacka:2012bm,Wilkinson:2001vv,Zumalacarregui:2017qqd,Oguri:2017ock,Macho:2000nvd,EROS-2:2006ryy,Griest:2013aaa,Griest:2013esa,Smyth:2019whb}, dynamical effects\cite{Graham:2015apa,Capela:2013yf,Capela:2012jz,Carr:2018rid,Monroy-Rodriguez:2014ula,MUSE:2020qbo,Koushiappas:2017chw}.
\vspace{0.5 em}

\textbf{PGWs induced density perturbations} We consider the spatially flat Friedmann-Lemaitre-Robertson-Walker (FLRW) spacetime perturbed up to second order with negligible anisotropic stress
 \begin{equation}
     ds^{2}=a(\eta)^{2} \,\left[-d\eta^{2} + \gamma_{ij}(x,\eta)\, dx^{i}dx^{j} \right],
 \end{equation}
 where $a(\eta)$ is the scale factor in conformal time and $\gamma_{ij}$ is the conformal spatial metric containing scalar and tensor perturbations at first and second order. At second order $\gamma_{ij}$ is defined as : $(1 - \phi^{(2)})\,\delta_{ij}  + \left(1/2\right) (\partial_{i}\partial_{j} - (1/3)\,\nabla^{2} \delta_{ij}) \, \chi^{||\,(2)} + \chi^{(1)\,T}_{ij}$ with $\chi^{(1)\,T}_{ij}$ being first-order tensor modes and $ \phi^{(2)}$ and $\chi^{||\,(2)}$ are tensor induced scalar perturbations. We assume the Universe is filled with a radiation fluid whose large overdensities collapse after the horizon entry, leading to the formation of PBHs. The conservation equations for second-order density $(\delta^{(2)})$ and velocity $(v^{(2)})$ perturbations are: 
 \begin{equation}
     \begin{split}
         {\delta^{(2)}}' + \frac{4}{3}\,\nabla^{2}\,v^{(2)} - 4\,{\phi^{(2)}}' & = \frac{4}{3} \, {\chi^{(1)}}^{ij} \,{\chi^{(1)}_{ij}}'\,,\\
         \delta^{(2)} + 4\,{v^{(2)}}' & = 0\,.
     \end{split} \label{eq:em_pert}
 \end{equation}
 Here, primes denote the derivatives with respect to conformal time, and $\mathcal{H}$ is the conformal Hubble parameter. The evolution of $\phi^{(2)}$ is governed by the Einstein equations (time-time, time-space, and space-space components) \cite{Bari:2022grh}. Linear GWs ($ {\chi_{ij}}^{(1)} \equiv \chi_{ij}$) evolve as:
 \begin{eqnarray}
     {\chi_{ij}}'' + 2 \,\mathcal{H}\, {\chi_{ij}}' - \nabla^{2}\, {\chi_{ij}} = 0\,.
 \end{eqnarray} 
The first-order tensor perturbation $\chi_{ij}(\textbf{x},\eta)$ is decomposed in Fourier space as
\begin{equation}
    \chi_{ij}(\textbf{x},\eta) =  \sum_{\sigma} \int \frac{d^{3} \bk}{\left(2 \pi\right)^{3/2}}\,e^{i \textbf{k}.\textbf{x}} \, \chi_{\sigma}\left(\bk,\eta\right)\,\epsilon_{ij}^{\sigma}\left(\hat{\bk}\right)\,, \label{eq:tensor_modes}
\end{equation}
where $\epsilon_{ij}^{\sigma}(\hat{\bk})$ are the two polarization tensors ($ \sigma = +,\times$) satisfying $\epsilon_{ij}^{\sigma}(\hat{\bk})\,{\epsilon^{\sigma' ij}}(\hat{\bk}) = 2 \delta_{\sigma \sigma'}$, and $\chi_{\sigma}(\bk,\eta)$ are the GW modes sourcing the density fluctuations. The temporal evolution of $\chi_{\sigma}\left(\bk,\eta\right)$ can be decomposed into its primordial value and a transfer function \cite{Picard:2023sbz,Bari:2022grh} as $\chi_{\sigma}(\bk)\,\sqrt{\pi / 2\,x} \, J_{1/2}(x)$, where $J_{1/2}(x\equiv k \eta)$ is the Bessel function of order $1/2$. The Fourier modes of density and velocity perturbations are also given in a similar fashion without the polarization tensors. 

Combining \cref{eq:em_pert} with Einstein equations yields the velocity fluctuation evolution in Fourier space \cite{Bari:2022grh}: 
\begin{equation}
{v^{(2)}}''' + \mathcal{H} \, {v^{(2)}}'' - \left[4 \mathcal{H}^{2} - \frac{1}{3} k^{2}\right] {v^{(2)}}'+ \frac{\mathcal{H}}{3} k^{2}{v^{(2)}}  = \mathcal{S}\,. \label{eq:TIV}
\end{equation}
Here, $\mathcal{S}(k,\eta)$ is the Fourier transform of $-\frac{1}{6}{\chi^{ij}}' {\chi_{ij}}'$ and $ v = v\left(k,\eta\right)$. It is important to note that the LHS of \cref{eq:TIV} reduces to the linear velocity perturbation evolution when the source term $\mathcal{S}$ vanishes. The solution of \cref{eq:TIV} is obtained, by transformation of ${v^{(2)}}$ to ${u^{(2)}} = \left(a\,{v^{(2)}} \right)' / a$, using the Green's function method. The tensor-induced density contrast $\delta^{(2)}(\eta,k)$ takes the following form:
\begin{equation}
\begin{split}
\delta^{(2)}(\eta,k) = -\frac{\pi}{12}\,\sum_{\sigma,\sigma'}\,\int \frac{d^{3} \textbf{q}}{\left(2 \pi \right)^{3/2}} \epsilon_{ij}^{\sigma}(\hat{\textbf{q}}) \, \epsilon^{\sigma' ij}(\widehat{\textbf{k}-\textbf{q}}) \\ \left(\frac{q}{k} \frac{|\textbf{k} -\textbf{q}|}{k}\right)^{-1/2}\,\mathcal{I}\left(x,q,|\textbf{k}-\textbf{q}|\right)\,\chi^{\sigma}(\textbf{k})\,\chi^{\sigma'}(\textbf{k}-\textbf{q}) \,.
\end{split} \label{eq:delta}
\end{equation}
The power spectrum of  density perturbations induced by primordial tensors follows from the equal-time two-point correlators defined as 
\begin{equation}
    \langle\delta^{(2)}(\eta,\bk)\, \delta^{(2)}(\eta,\bk') \rangle = \delta^{(3)} \left(\bk + \bk'\right)\,\frac{2\pi^{2}}{k^{3}} \mathcal{P}_{\delta^{(2)}}\left(\eta,k \right)\,. \label{eq:PSD}
\end{equation}
Using \cref{eq:delta,eq:PSD}, the dimensionless power spectrum $\mathcal{P}_{\delta^{(2)}}\left(\eta,k \right)$ is computed as following:
\begin{equation}
\mathcal{P}_{\delta^{(2)}}= \frac{1}{2}\int_{0}^{\infty}dv \int_{\abs{1-v}}^{1+v} \frac{f(u,v)}{\left(u v \right)^{3}}\overline{\mathcal{I}^{2}\left(u,v\right)} \mathcal{P}_{\chi}(ku) \mathcal{P}_{\chi}(kv) \,. \label{eq:PSDf}
\end{equation}
Here $f(u,v)$ contracts the polarization tensors \cite{Bari:2021xvf}. We employ a change of variables namely $v = q/k$, $ u = |\bk -\bf{q}|/\bk$ and $x = k\,\eta$ to arrive at the above form of $\mathcal{P}_{\delta^{(2)}}\left(\eta,k \right)$ in \cref{eq:PSDf}. The kernel $\mathcal{I}(x,u,v)$ constitutes the following time integrals of the product of three Bessel functions 
\begin{equation}
\begin{split}
    \mathcal{I}_{a}\left(x,u,v\right) =\int_{0}^{x}d\bar{x} \,\bar{x}^{-1/2} J_{3/2}\left(\frac{\bar{x}}{\sqrt{3}}\right) J_{3/2}\left(v \bar{x}\right) J_{3/2}\left(u\bar{x} \right) \,,\\
    \mathcal{I}_{b}\left(x,u,v\right) = \int_{0}^{x}d\bar{x} \,\bar{x}^{-1/2} Y_{3/2}\left(\frac{\bar{x}}{\sqrt{3}}\right) J_{3/2}\left(v \bar{x}\right) J_{3/2}\left(u\bar{x} \right)\,.
\end{split}   \label{eq:kernelab}
\end{equation}
These integrals should be evaluated for any values of $x$ but the exact form of \cref{eq:kernelab} is analytically obscure without giving a clear idea of time dependence. However, we can obtain the analytical formula for the integrals for the scales leaving  sub-horizon $(x \gg1)$ by extrapolating the upper limits of integrals to infinity\cite{threebesselI,NIST:DLMF,originalbessel}. We have explicitly checked this approximation using the numerical integral methods. The kernel $\mathcal{I}(x,u,v)$ is defined as $F_{1}(x) \mathcal{I}_{a}(u,v) + F_{2}(x) \mathcal{I}_{b}(u,v)$, where $F_1(x)$ and $F_2(x)$ are oscillatory functions composed of Bessel and Hypergeometric functions. The final form of time-average of kernel becomes $   \overline{\mathcal{I}^{2}(u,v)}  = \frac{\pi^{3}}{144}\, \left[{\mathcal{I}_{a}^{2}}(u,v) +  \mathcal{I}_{b}^{2}(u,v)\right]$, with \cite{Espinosa:2018eve} 
\begin{equation}
    \begin{split}
        \mathcal{I}_{a}(u,v) = \frac{3^{3/4}}{2} \,\sqrt{\frac{2\,u v}{\pi}}\,\left(1 - y^{2}\right)\,\Theta \left(u + v - \frac{1}{\sqrt{3}}\right) \,,\\
        \mathcal{I}_{b}(u,v) = \frac{3^{3/4}}{2\pi}\, \sqrt{\frac{2\,u v}{\pi}} \, \left[ y +\frac{\left(1 - y^{2}\right)}{2}\, \log\left(\frac{1+y}{1-y}\right)\right] \,,
\end{split} \label{eq:anakernelab}
\end{equation}
where $y = \left(u^2 + v^2 -1/3\right) / \left(2 u v\right)$. As shown in \cref{eq:PSDf}, the primordial tensor power spectrum influences the spectral shape of tensor-induced density perturbations irrespective of the GW source. In this work, we restrict ourselves to PGWs only. This primordial tensor power spectrum not only sources the second-order density power spectrum (leading to PBH formation) but also generates the SGWB observed by NANOGrav and other PTAs. In the following, we consider several ansatze for the tensor power spectrum, each with a distinct spectral shape applicable to different classes of models.

We first consider the typical power-law spectrum with pivot scale $k_s = 0.05 \rm Mpc^{-1}$ predicted by single-field inflationary models \cite{Gong:2014qga,Mukohyama:2014gba,Kuroyanagi:2014nba}
\begin{equation}
    \mathcal{P}^{\rm PL}_{h}(k) = r\,A_{s}\,\left(\frac{k}{k_{s}}\right)^{n_T}  \, \Theta(k_{\rm max}-k)\,. \label{eq:PL}
\end{equation}
where $r$ is the tensor-to-scalar ratio with an upper bound $r\leq0.032$ inferred from the joint analysis of Planck and BICEP/Keck \cite{BICEP:2021xfz}, $A_s = 2.1 \times 10^{-9}$ is the amplitude of curvature perturbations measured at $k_s$\cite{Planck:2018vyg}, and $n_t$ denotes the spectral tilt. Another class of spectra is characterized by a log-normal(LN) shape predicted by hybrid inflation and curvaton models \cite{Kawasaki:2012wr,Frampton:2010sw,Pi:2017gih}
\begin{equation}
    \mathcal{P}^{\rm LN}_{h}(k) = \frac{A_{T}}{\sqrt{2\,\pi}\,\Delta}\, \exp\left(-\frac{1}{2 \Delta^{2}} \ln^{2}\left(k/k_{p}\right)\right) \,, \label{eq:LN}
\end{equation}
where $A_{\rm T}$ is amplitude, $k_{p}$ is peak of frequency, and $\Delta$ is width of LN spectrum. We present a concrete example of an inflationary model relating the $A_{\rm T}$ and $\Delta$ to the inflationary model parameters in \textit{supplementary material}. The present-day spectrum of GW energy density fraction is given by \cite{Caprini:2018mtu}
\begin{eqnarray}
    h^2 \Omega_{\rm GW}(k) =\frac{\Omega_{r}^{0} h^2}{24} \mathcal{P}_{h}(k)\,,
\end{eqnarray}
where $\Omega_{r}^{0} h^2 \simeq 4.2 \times 10^{-5}$ is the current energy density. The current GW frequency corresponds to $ k = 6.25 \times 10^{5} \, \textrm{Mpc}^{-1} \left( f / \textrm{nHz}\right)$. Furthermore, the GW energy spectrum contributes to the effective number of relativistic degrees of freedom parameterized by $N_{\rm eff}$ defined as\cite{Caprini:2018mtu,Boyle:2007zx,Allen:1997ad,Smith:2006nka,Kuroyanagi:2014nba,Ben-Dayan:2019gll} 
\begin{equation}
    \int_{f_{\rm BBN}}^{f_{\rm max}} \frac{df}{f}\, \Omega_{\rm GW}(f) h^2 \leq 5.6 \times 10^{-6} \, \Delta N_{\rm eff}\,. \label{eq:Neff}
\end{equation}
Here, $\Delta N_{\rm eff}$ represents the deviation from its standard model value. The frequency corresponding to the time of the BBN era is $f_{\rm BBN} = 1.84 \times 10^{-11} $ Hz. The upper limit of \cref{eq:Neff} is then fixed so that observational bounds on $\Delta N_{\rm eff} $ are not violated. The 2 $\sigma$ upper limit on $\Delta N_{\rm eff}$ is constrained by BBN and CMB probes to be $\Delta N_{\rm eff} \leq 0.4$.\cite{Planck:2018vyg,Aver:2015iza,Cooke:2017cwo,Vagnozzi:2019ezj,2020ApJ...896...77H,ACT:2020gnv,Mossa:2020gjc}.
\begin{figure*}[ht]
\includegraphics[width=8.5 cm,height=6.5 cm]{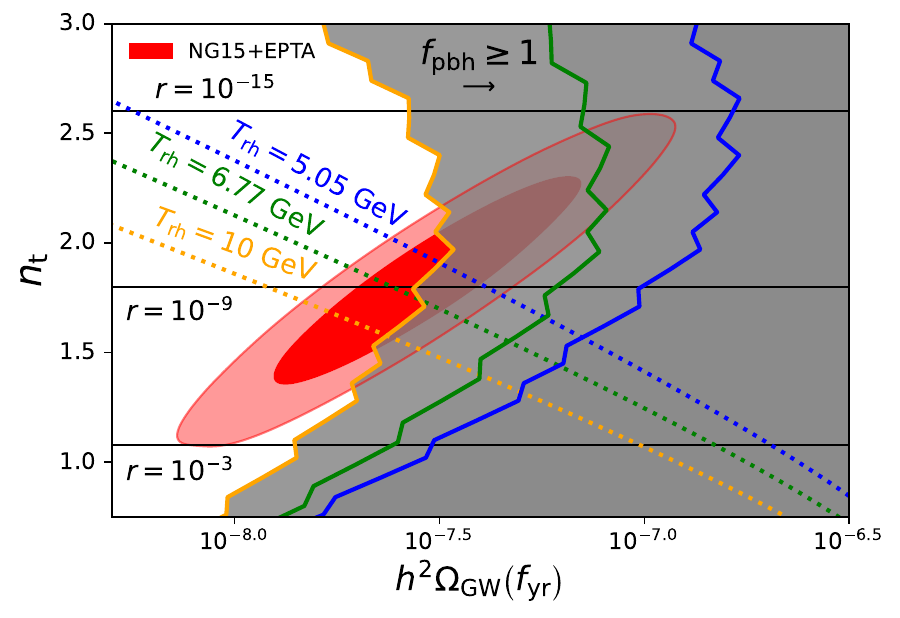}
\includegraphics[width=8.5 cm,height=6.5 cm]{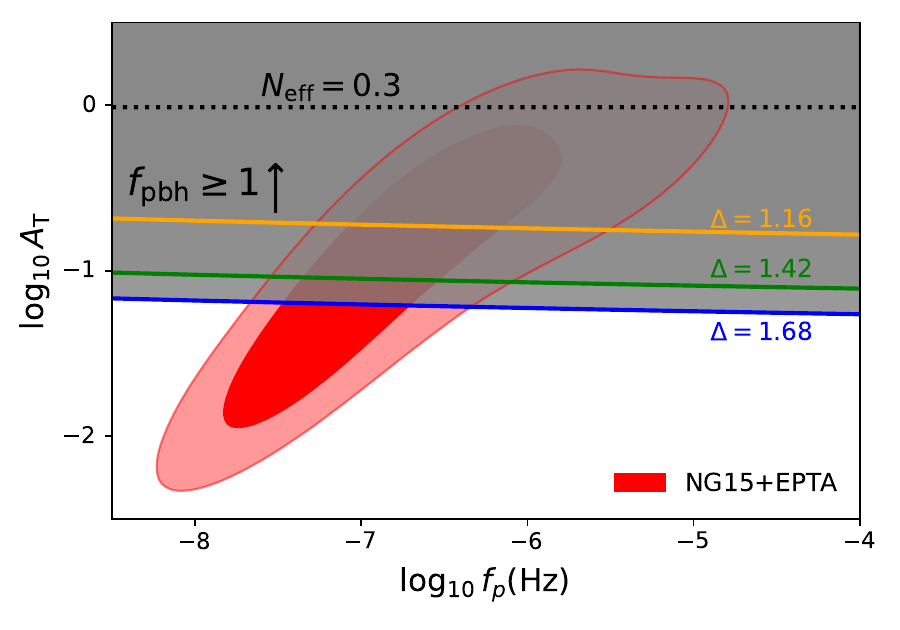}
\caption{\textit{Left panel:} Two dimensional contour plot for inferred posterior values of $n_t$ and present day GW energy density evaluated at reference frequency $f_{\rm yr}$ obtained from the joint analysis of NG-15 and EPTA-DR2 datasets. Black curves correspond to the tensor-to-scalar ratio as shown. We include the allowed values of reheating temperatures that satisfy the $N_{\rm eff}$ bound (shown in dotted lines) and the corresponding viable parameter space (on the left) to avoid the overproduction of PBHs, given that $f_{\rm PBH} \geq 1$. \textit{Right panel}: same as left panel but for the LN model. Again, $N_{\rm eff} = 0.3$ constraint given in black dashed line. The parameter space lying below blue ($\Delta$ = 1.68), green ($\Delta$ = 1.42), and orange ($\Delta$ = 1.16) respectively survives the $f_{\rm pbh} \geq 1$.} \label{fig:PLEN}
\end{figure*}
\begin{table}[ht]
\centering
\setlength{\tabcolsep}{1.0pt}
\renewcommand{\arraystretch}{1.3}
\begin{tabular}{|p{0.5cm}|p{1.35cm}|p{1.35cm}|c|c|c|}
\hline
\hline
\multirow{2}{*}{\textbf{}}  & \multirow{2}{*}{\textbf{Param}} & \multirow{2}{*}{\textbf{Prior}} & \multirow{2}{*}{\textbf{NG-15}} & \multirow{2}{*}{\textbf{EPTA2}} & \textbf{NG-15} \\
      & &  &   &  & \textbf{+EPTA2} \\
\hline
\hline
\multirow{2}{*}{PL} 
    & $\log_{10} r$ & $[-40,-1]$ & $-9.5 \pm 2.8 $  & $-8.1^{+5.6}_{-2.5} $ & $-8.8\pm 2.5$ \\
    & $n_t$          & $[0,4]$   & $1.91\pm 0.36$  & $1.73^{+0.31}_{-0.71} $ & $1.83\pm 0.31$\\
\hline
\multirow{3}{*}{LN} 
    & $\log_{10} A_{T}$ & $[-10,0]$  & $-0.88^{+0.67}_{-0.46}$ & $-1.33^{+0.85}_{-0.66}$ & $-0.93^{+0.68}_{-0.46} $\\
    & $\Delta$          & $ [0.01,4]$   & $1.42^{+0.26}_{-0.53}$ & $2.7^{+1.1}_{-1.4}$ & $1.44^{+0.24}_{-0.54}$\\
    & $\log_{10} f_p$   & $[-10,-2]$  & $-6.73^{+0.54}_{-0.76}$ & $-6.2^{+1.4}_{-1.6}$ & $-6.73^{+0.37}_{-0.94}$\\
\hline
\hline
\end{tabular} 
\caption{Prior and mean $\pm 1 \sigma$ constraints for power-law (PL) and log-normal (LN) spectra for NANOGrav-15, EPTA DR2, and their combination.}
\label{tab:models}
\end{table}

\textbf{PBH formation:} The PBHs are formed when the gravitational collapse of density perturbation $\delta(\eta,\bk)$ exceeds a critical threshold $\delta_c$. Since the first-order tensor and first-order scalar modes are statistically independent, the total density contrast $\delta(\eta,\bk)$  upto second order is computed by simply adding both contributions as 
\begin{equation} 
    \delta(\eta,\bk) = \frac{4}{9} \, \frac{k^2}{\mathcal{H}^2} \, \mathcal{T}(\eta,\bk)\,\zeta^{(1)}(\bk) + \frac{1}{2}\,\delta^{(2)}(\eta,\bk) \,.\label{eq:total+density}
\end{equation} 
Here we have related the first-order density perturbation $\delta^{(1)}(\eta,\bk)$ to the first-order primordial curvature perturbation $\zeta^{(1)}({\bk}) $ via a transfer function $\mathcal{T}(\eta,\bk)$. Given that the first-order curvature perturbations are scale-invariant and their amplitude is tightly constrained at large scales using the Planck's CMB measurements, the contribution of linear modes in \cref{eq:total+density} is sub-dominant given that curvature perturbations remain scale-invariant at small scales as well. However tensor-induced density perturbations start to dominate at small scales. We adopt the Press–Schechter formalism \cite{Farooq:2025vmp,Green:2004wb} to compute the fraction of the density collapse to form PBHs, $\beta$ is given by
\begin{eqnarray}
    \beta(M) = \int_{\delta_{c}}^{\infty} \, P(\delta) \, d \delta \equiv \frac{1}{2}\, \text{erfc}\left\{ \frac{\delta_c}{\sqrt{2}\,\sigma_{\delta}(M)}\right\}\,, \label{beta}
\end{eqnarray} 
\begin{figure*}[ht]
\includegraphics[width=8.8 cm,height=7.0 cm]{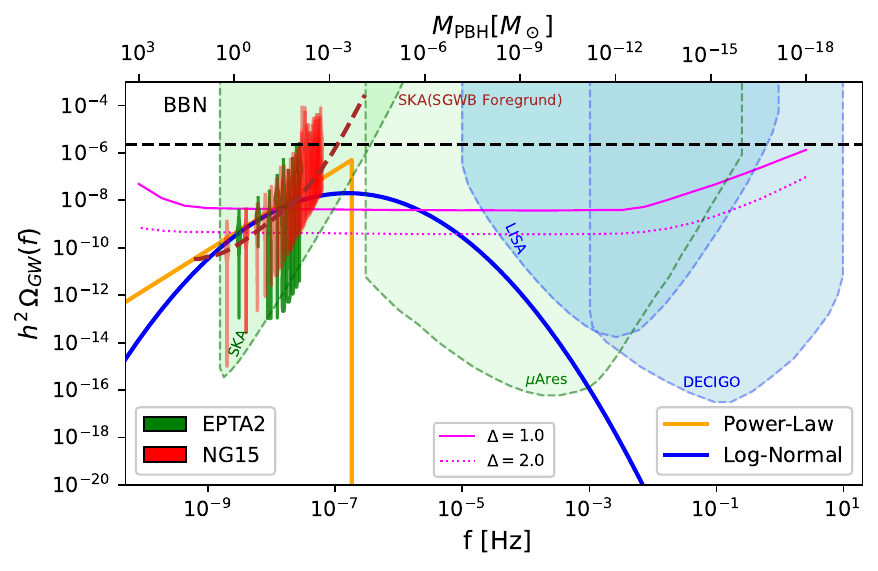}
\includegraphics[width=8.8 cm,height=7.0 cm]{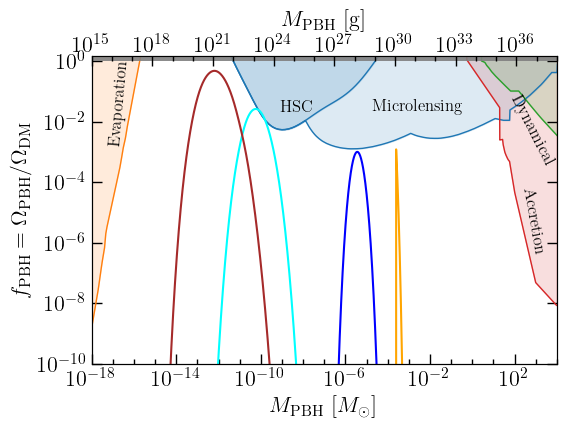}
\caption{\textit{Left panel:} GW power spectra obtained from PL and LN spectra in orange and blue respectively. Shown are the violin plots in green and red for EPTA2 and NG15 measurements respectively, and the sensitivity curves for SKA, $\mu$Ares, LISA, and DECIGO in colors indicated. We also have included the updated SKA sensitivity curve using the SGWB as a foreground \cite{Babak:2024yhu,Cecchini:2025oks} in brown. \textit{Right panel:} Mass function resulting from the PL (orange) and LN spectrum for the values of parameters described in the text. The PBHs abundance constrained by variety of observations including PBH evaporation\cite{Katz:2018zrn,Montero-Camacho:2019jte,Arbey:2019vqx,Boudaud:2018hqb,DeRocco:2019fjq,Laha:2020ivk,Carr:2009jm,Ballesteros:2019exr,Laha:2019ssq,Poulter:2019ooo,Katz:2018zrn,Saha:2024ies,Saha:2021pqf}, gravitational lensing\cite{Niikura:2019kqi,Niikura:2017zjd,Barnacka:2012bm,Wilkinson:2001vv,Zumalacarregui:2017qqd,Oguri:2017ock,Macho:2000nvd,EROS-2:2006ryy,Griest:2013aaa,Griest:2013esa,Smyth:2019whb}, dynamical effects\cite{Graham:2015apa,Capela:2013yf,Capela:2012jz,Carr:2018rid,Monroy-Rodriguez:2014ula,MUSE:2020qbo,Koushiappas:2017chw} are also shown. We have chosen $\delta_c = 0.42$ for our analysis.}  \label{fig:fpbh}
\end{figure*}
The above relation assumes that the probability distribution of density perturbation, $P(\delta)$, is Gaussian. In principle, the induced density contrast are highly non-Gaussian due to its quadratic dependence on the primordial tensor perturbation \cite{Abdelaziz:2025qpn}. While including non-Gaussianity (NG) would further improve the results, this paper focuses on demonstrating the viability of tensor-induced density perturbations for PBH formation. A comprehensive analysis incorporating ineludible NGs is deferred to future work. The variance in the density fluctuations smoothed over a radius $R$ using the Gaussian window function \cite{Tokeshi:2020tjq} $W(k,R) = \text{exp}\left(-k^2 R^2 / 2 \right)$ ,denoted as $\sigma^2_{\delta}$, obtained from 
\begin{eqnarray}
    \sigma^2_{\delta}(M) = \int_{0}^{\infty} \, \frac{dk}{k} \mathcal{P}_{\delta}(k,\eta=R)\,W^{2}(k,R) \,,\label{eq:variance} 
\end{eqnarray}
where $M$ is the PBH mass and $R  = 1 / k(M)$ is smoothing scale for the PBH formation. The current fraction of PBHs as DM over a logarithmic interval is 
\begin{equation}
    f_{\rm PBH}(M) = \frac{1}{\Omega_{\rm DM}} \frac{d\,\Omega_{\rm DM}}{d\, \ln M} \simeq \frac{\beta(M)}{1.84 \times 10^{-8}}\, \left(\frac{M}{M_{\odot}}\right)^{-1/2}\,.
\end{equation}
Finally, the mass of PBH is related to the scale of perturbation for PBH production as \cite{Inomata:2018cht}
\begin{equation}
    M_{\rm PBH} \simeq   M_{\odot} \left(\frac{\gamma}{0.2}\right) \left(\frac{g}{10.75}\right)^{-\frac{1}{6}} \, \left(\frac{k}{1.9 \times 10^{5} \text{Mpc}^{-1}}\right)^{-2}\,.
\end{equation}
where $\gamma = 0.2$ is the fraction of PBH mass in the horizon mass at the production \cite{Carr:1975qj}, and $g = 10.75$ is the effective degrees of freedom for radiation at that time. We fix $\delta_c = 0.42$ for the computation of PBH abundance \cite{Carr:2020xqk}.

\vspace{0.5 em}
\textbf{Analysis and results:}
We perform a joint Bayesian analysis of the NANOGrav15 (NG15) and EPTA DR2 (EPTA2) datasets. The joint constraints are derived by reweighting the NG15 chain samples, generated using \texttt{PTArcade}\cite{Mitridate:2023oar}, based on posterior densities estimated from the EPTA2 chains via Gaussian Kernel Density Estimation. We then resample the full NG15 chain according to these computed weights. We use the publicly available tool \texttt{GetDist}\cite{Lewis:2019xzd} for analyzing chains. The prior information and mean$\pm 1 \sigma$ constraints for both models are given in \cref{tab:models}. The two-dimensional contour plots for the inferred posterior values for PL and LN scenarios are shown in the left and right panels of \cref{fig:PLEN} respectively. The dashed curves in the left panel of \cref{fig:PLEN} represent contours of $\Delta N_{\rm eff} = 0.4$ for different values of the reheating temperature $T_{\rm rh}$. These contours consequently determine the $k_{\rm max}$ values appearing in \cref{eq:PL,eq:Neff}. The inference of blue spectrum and corresponding reheating temperatures required to fulfill $\Delta N_{\rm eff}$ bound also has interesting consequences for formation of PBHs induced from the tensor modes. The blue spectral index of the PL spectrum provides the rise of the tensor-induced density power spectrum while reheating temperature fixes its peak (\textit{see supplementary material}). The solid curves show the constraints on the spectral index ($n_t$) and $h^2 \Omega_{\rm GW}(f_{\rm yr})$ avoiding the overproduction of PBHs. The PL GW interpretation of the PTA signal in red survives but cannot account for the dark matter. The each isocontours of $f_{\rm pbh} = 1$ corresponds to same $f_{\max}$ allowed to satisfy the $N_{\rm eff}$ constraints. It is important to note that for reheating temperatures as low as $5.05$ GeV the whole parameter space is compatible with the PTA observations and also avoids the overproduction of the PBHs. It is striking to see from the \cref{tab:models} (\textit{and also from full posterior distributions in supplementary material}) that NG15 offers stringent constraints compared to EPTA2 for LN specifically. We again show the allowed values of $A_{T}$ for different values of $\Delta$ spread over the mean $ \pm 1\sigma$ constraints satisfying $f_{\rm pbh} = 1$ and $N_{\rm eff} = 0.3$. See the SM for the evolution of the density power spectrum for the LN scenario.

The left panel of \cref{fig:fpbh} displays the gravitational wave spectra for power-law (PL, orange) and log-normal (LN, blue) models, using parameters within the $1\sigma$ confidence region from joint PTA analyses. For the PL spectrum, we adopt $r = 2.56 \times 10^{-8}$, $n_t = 1.68$, and $k_{\mathrm{max}} = 1.2 \times 10^{8}~\mathrm{Mpc}^{-1}$ (corresponding to $T_{\mathrm{rh}} = 10~\mathrm{GeV}$). The LN spectrum uses $A_T = 9.2 \times 10^{-2}$, $\Delta = 1.42$, and $\log_{10} f_{p} = -6.8$. These parameters yield PBH abundances of $10^{-3} M_{\odot}$ and $10^{-5} M_{\odot}$ for the PL and LN models, respectively. Tight constraints on $k_{\mathrm{max}}$ from $\Delta N_{\mathrm{eff}}$ measurements limit the PL spectrum to PBH masses near $10^{-3} M_{\odot}$ while simultaneously explaining both NG15 and EPTA datasets. In contrast, the LN model's PBH mass range is determined by the peak frequency $f_p$, allowing broader mass possibilities. As mentioned earlier, the EPTA2 has a wide parameter space that can account for the PBHs as dark matter. The mass function resulting from the $95 \%$ C.I. posterior values of the EPTA2 parameter space in cyan and brown color in right panel of \cref{fig:fpbh}. We overlay the corresponding parameter values in the same color in SM (see right panel, \cref{fig:triangle_PL}).

\textbf{Constraints on $h^2 \, \Omega_{\rm GW}$ from $f_{\rm PBH}$:} \cref{fig:PLEN,fig:fpbh} show that the $f_{\rm pbh}$ imposes additional constraints on inferred model parameters. The abundance of PBHs has been constrained by a variety of observations, including extra-galactic $\gamma$-rays from PBH evaporation\cite{Katz:2018zrn,Montero-Camacho:2019jte,Arbey:2019vqx,Boudaud:2018hqb,DeRocco:2019fjq,Laha:2020ivk,Carr:2009jm,Ballesteros:2019exr,Laha:2019ssq,Poulter:2019ooo}, femtolensing of $\gamma$-ray bursts \cite{Katz:2018zrn}, Subaru/HSC microlensing\cite{Dasgupta:2019cae,Laha:2020vhg}, Kepler milli/microlensing \cite{Griest:2013esa}, OGLE microlensing \cite{Niikura:2019kqi}, EROS/MACHO microlensing\cite{EROS-2:2006ryy}. Using these observational bounds, we derive the amplitude of the SGWB when interpreted as PGWs. For the LN spectrum, we compute the allowed values of $A_{\rm T}$ and consequently $h^2\Omega_{\rm GW}$ at each peak frequency $f_p$ consistent with $f_{\rm pbh}$ constraints. This analysis considers spectral width parameters $\Delta = 1.0$ and $\Delta = 2.0$, with the resulting SGWB spectra presented in the left panel of \cref{fig:fpbh}. Solid and dotted orange lines respectively denote the $\Delta = 1.0$ and $\Delta = 2.0$ cases. It is interesting to note that the derived $\Omega_{\rm GW}(f_p)$ amplitudes fall within the sensitivity ranges of several next-generation gravitational wave observatories, indicating the inevitability of the production of PBHs from the PGWs.

\textbf{Effects of NGs of PBH abundance:} As shown in the right panel of \cref{fig:fpbh}, the peak of the PBH mass function is determined by $f_{\rm max}$ for the PL spectrum and $f_p$ for the LN spectrum, respectively. We expect ineludible NG effects not to alter this peak location, though their inclusion can expand the allowed parameter space for tensor mode amplitudes and prevent PBH overproduction. In general, NG significantly enhances or suppresses the tail of $P(\delta)$ while leaving the variance unchanged. Therefore, for a fixed PBH abundance, our mechanism’s predictions remain broadly robust to NG effects \cite{Pi:2024lsu}. Recently, the Bispectrum quantifying NG in tensor-induced matter density perturbations was computed in \cite{Abdelaziz:2025qpn}. This work shows that the NG amplitude depends on the tensor power spectrum’s shape, with the LN spectrum generating substantially larger NGs than the PL spectrum. Although we consider radiation density perturbations (not matter perturbations), similar conclusions likely apply in our case. Thus, we expect our PL spectrum results to remain largely unchanged, whereas the LN spectrum results could differ significantly. We leave the computation of the Bispectrum and NG probability density function for this scenario to future work.

\textbf{Conclusions and outlook:} In this paper, we have analyzed a novel mechanism for PBH formation from the collapse of radiation overdensities induced by first-order PGWs. PBH production in such a scenario occurs irrespective of the PGW source. Using Inflationary PGWs as a potential explanation for the observed SGWB, we performed a joint log-likelihood analysis with NANOGrav and EPTA2 data. Our results demonstrate that PGWs can generate a sizable abundance of PBHs while simultaneously satisfying constraints from PBH overproduction and the NG15+EPTA2 datasets. Furthermore, our formalism provides a framework for assessing the detectability prospects of the PGW background, ensuring consistency with observational bounds on the PBH fraction ($f_{\rm pbh}$) across a wide range of scales. We explicitly show that future missions, including SKA, LISA, and DECIGO, could produce PBHs acting as primordial seeds for supermassive black holes or potentially explain dark matter partially or entirely.

The present analysis can be extended in several ways. It has been shown in several previous works that computation of PBH abundance comes with many levels of uncertainties including the underlying theory, value of critical threshold, and choices of different types of window functions \cite{Ando:2018qdb,Young:2019osy,Musco:2023dak}. Future work could employ peak theory to calculate PBH abundance more precisely \cite{Young:2022phe,Young:2020xmk,Young:2019yug}. Moreover, we omitted contributions from the ineludible non-Gaussianity arising from quadratic tensor modes in the source term. Incorporating this effect, along with considering the inherent non-Gaussian statistics of PGWs, would be valuable\cite{DeLuca:2019qsy,Abdelaziz:2025qpn,Shiraishi:2019yux,Peng:2024ktq,Cai:2018dig,Pi:2024lsu,Pi:2020otn}. We expect a comprehensive analysis accounting for these considerations will further refine the parameter space inferred from current PTA results and enhance the detection prospects for upcoming missions.

\textit{\textbf{Acknowledgments}}-- I thank Pritha Bari, Anish Ghoshal, Shi Pi, and Gabriele Franciolini for their useful comments.

 \newpage
\bibliography{ref.bib}
\clearpage
\onecolumngrid

\renewcommand{\thepage}{S\arabic{page}}
\renewcommand{\theequation}{S.\arabic{equation}}
\renewcommand{\thetable}{S.\Roman{table}}
\renewcommand{\thefigure}{S.\arabic{figure}}
\setcounter{page}{1}
\setcounter{equation}{0}
\setcounter{table}{0}
\setcounter{figure}{0}


\begin{center}
\textit{\Large Supplemental Material
}
\end{center}
\noindent

This supplemental material contains more details on the evolution of total density power spectrum and full constraints on model parameters.
\section{Evolution of density power spectrum}  The total density contrast $\delta$ up to the second order is given as 
    \begin{align}
       \begin{split}
           \delta(k,\eta) &=  \delta_{\bk}^{(1)}(x) \,+\,\frac{1}{2} \, \delta^{(2)}_{\bk}(x)  \equiv \frac{4}{9} \frac{k^2}{\mathcal{H}^{2}} \, T(x)\,\zeta^{(1)}_{k} \, \, + \,  \frac{1}{2} \,\delta^{(2)}_{\bk}(x) 
       \end{split} 
    \end{align} 
It is straightforward to derive the power spectrum by calculating the 2-point correlators as defined in main text. The total power spectrum of density perturbations is expressed as following
\begin{equation}
    \mathcal{P}_{\delta} (k,\eta) = \frac{16}{81}\,\left(\frac{k}{\mathcal{H}}\right)^{4}\, \mathcal{P}_{\zeta} + \frac{1}{4} \,\mathcal{P}_{\delta^{(2)}}(k,\eta) \equiv \frac{16}{81}\,A_{s} \,\left(\frac{k}{\mathcal{H}}\right)^{4}\,\left(\frac{k}{k_{s}}\right)^{n_{s} -1 }\, + \frac{1}{4} \,\mathcal{P}_{\delta^{(2)}}(k,\eta) \label{eq:totalPSD}
\end{equation}
\subsection{Power-Law Spectrum:} For the total power-law spectrum, the second-order density power spectrum is simplified as following:
\begin{equation}
    \mathcal{P}^{\rm PL}_{\delta^{(2)}}(k) \simeq \frac{1}{2}\,A_{s}^{2}\,r^{2} \, \left(\frac{k}{k_s}\right)^{2\,n_t}\int_{0}^{\infty} \, \int_{|1-v|}^{1+v} \, f(u,v) \, \overline{\mathcal{I}^{2}\left(u,v\right)} \,\left(u\,v\right)^{n_t-3} \equiv  A_{s}^{2} \, \mathcal{Q}\left(n_t,r\right)\,\left(\frac{k}{k_{s}}\right)^{2\,n_t} \label{eq:anaPLPSD}
\end{equation}

\begin{table}[ht]
\centering
\large 
\setlength{\tabcolsep}{12pt} 
\begin{tabular}{||c|c|c|c|c|c||}
\hline
\hline
\multicolumn{1}{||c|}{\textbf{\( n_t \)}} & 1.0 & 1.375 & 1.75 & 2.125 &  2.5\\
\hline
\multicolumn{1}{||c|}{\textbf{\( r \)}} & $ 10^{-3}$ & $10^{-6}$&$10^{-9}$ & $1 \times 10^{-12}$ & $10^{-15}$\\
\hline
\( \mathcal{Q}(n_t,r) \) & 385.15 & 193.03 & 8587.20 & 824122.921 & 91501961.32 \\
\hline
\hline
\end{tabular}
\caption{The coefficient $Q(n_t,r)$ estimated from the numerical integral for different values of $n_t$ and $r$.}
\end{table}
\begin{figure}[ht]
\includegraphics[width=8.5 cm,height=6.5 cm]{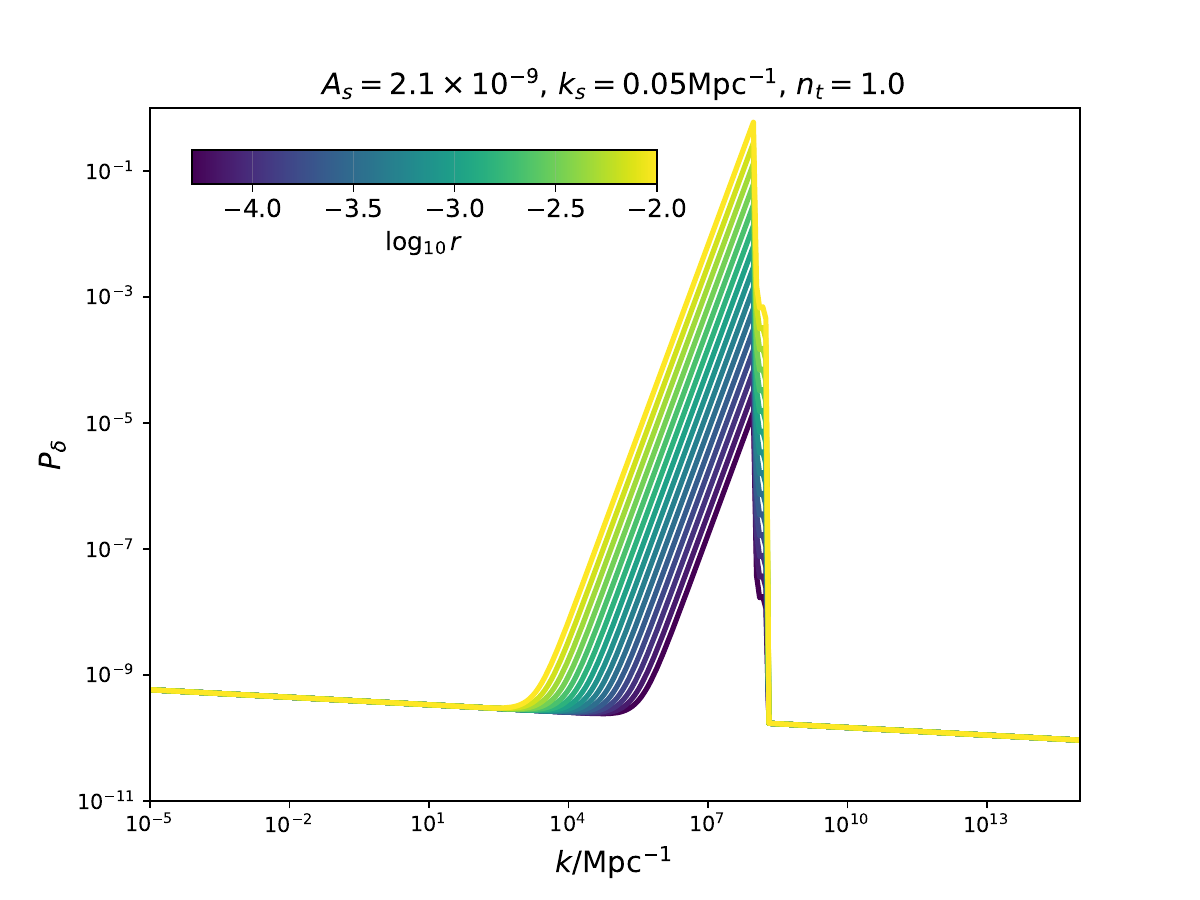}
\includegraphics[width=8.5 cm,height=6.5 cm]{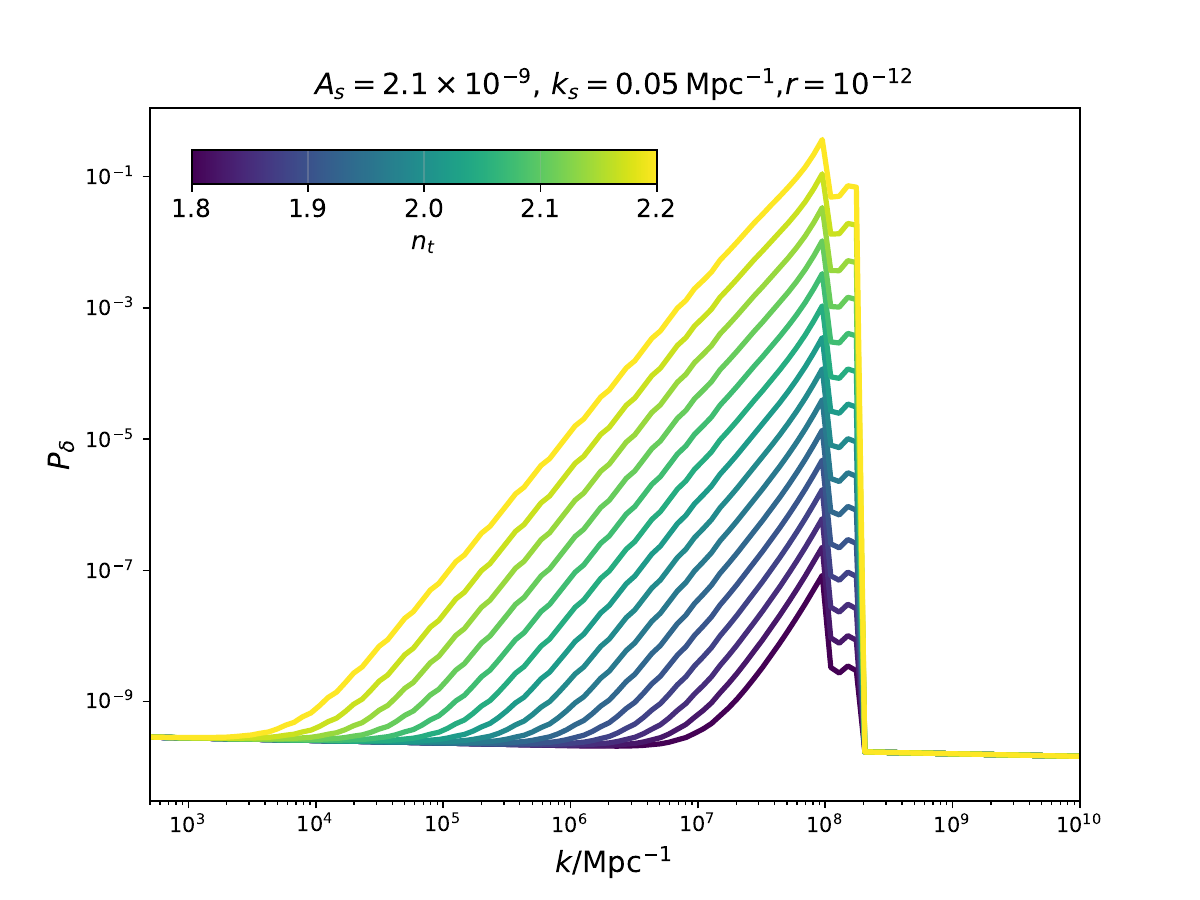}
\caption{Evolution of total density power spectrum for Power-law type spectrum. We have numerically integrated \cref{eq:anaPLPSD} using the expression given in \cref{eq:PL} (see main text). Due to the Heaviside function, we note that slope of the resulting expression differs from exact result slighting around its peak. We have fixed the scalar power spectrum parameters $A_s = 2.1 \times 10^{-9}\,,\, k_s = 0.05 \,\text{Mpc}^{-1}\,,$ and $n_s = 0.96$ according to CMB measurements \cite{Planck:2018vyg}. We fix the tensor spectral index $n_t = 1.0$ and vary to tensor-to-scalar ration $r \in [10^{-5}\,,\,10^{-2}\,]$ in left panel and vary $n_t \in [1.8\,,\,2.2]$ and $r = 10^{-12}$ in right panel.} \label{fig:secPSD}
\end{figure}
\begin{figure}[ht]
\includegraphics[width=8.5 cm,height=6.5 cm]{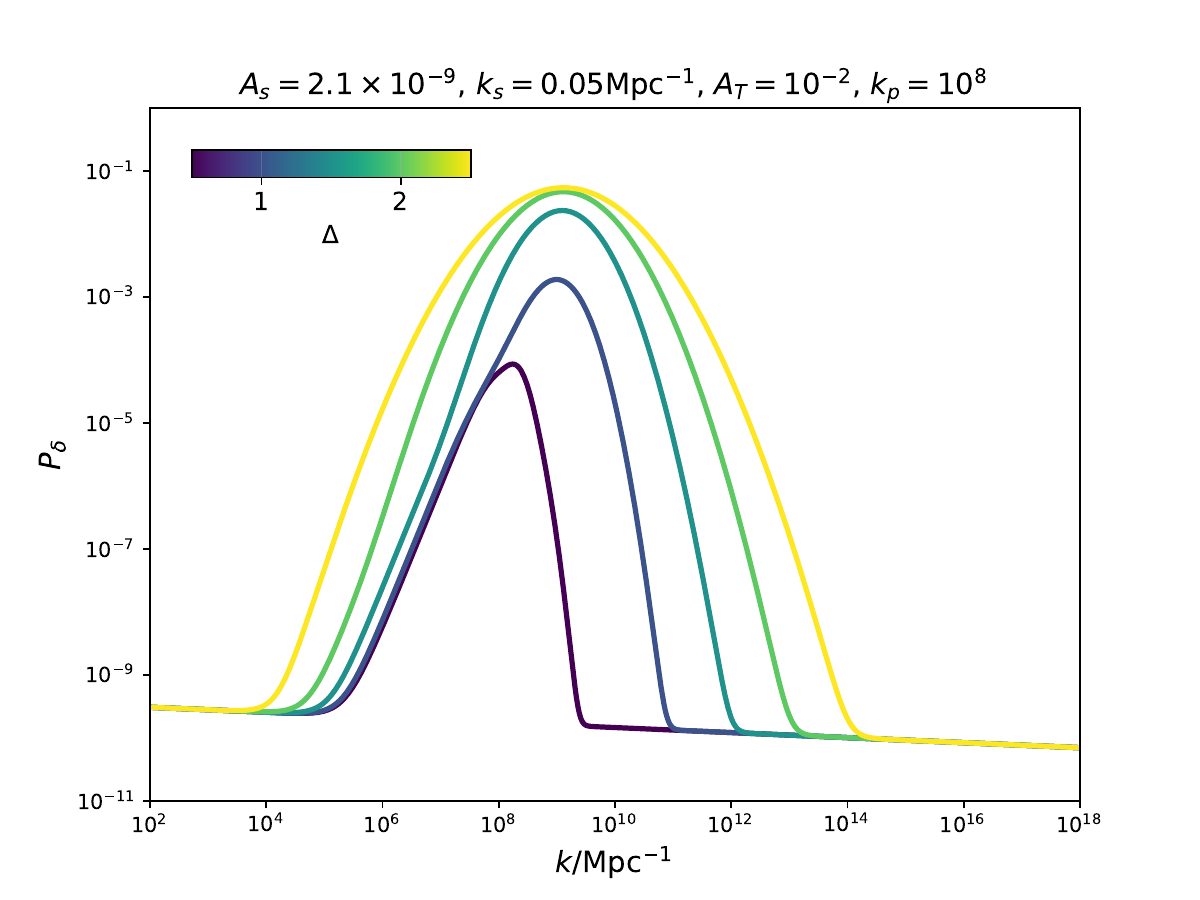}
\includegraphics[width=8.5 cm,height=6.5 cm]{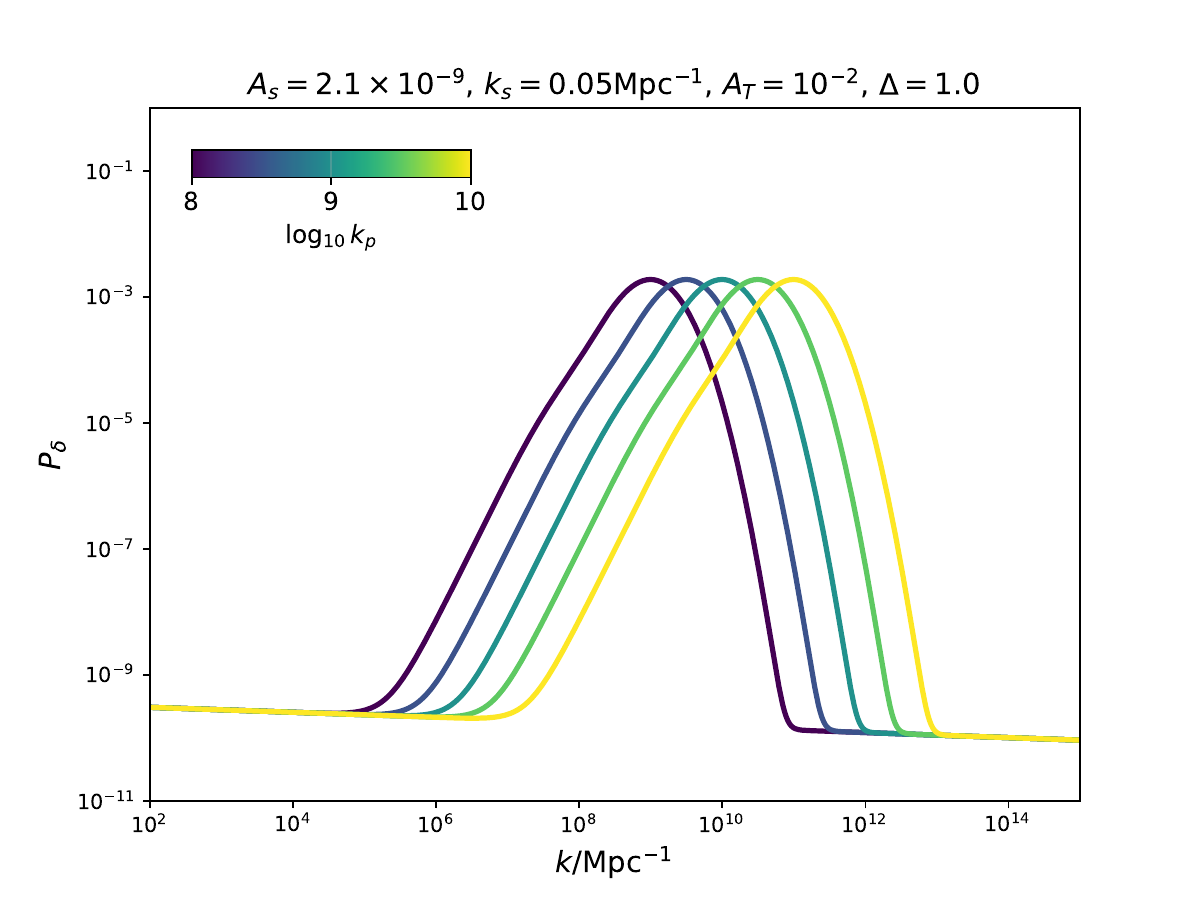}
\caption{Evolution of total density power spectrum for Log-Normal type spectrum. We show the impact of the spectrum width $\Delta$ and peak frequency $k_p$  on the total power spectrum in left and right panel respectively. Note that for $\Delta = 0.5$ the spectrum shows the similar slope as in \cref{eq:IRLN}.} \label{fig:secPSDLN}
\end{figure}
\subsection{Log-Normal Spectrum:} For the Log-Normal spectrum, the analytical solution can be found for $\Delta \rightarrow 0$. In this limit, LN power spectrum reduces to Dirac-Delta power spectrum and its analytical solution can be expressed as:
\begin{equation}
    \mathcal{P}^{\rm LN}_{\delta^{(2)}}(k) \overset{\Delta \rightarrow 0}{\simeq} \frac{1}{2} \, \left(\frac{k}{k_p}\right)^{4}\, f\left(\frac{k_p}{k},\frac{k_p}{k}\right) \, \overline{\mathcal{I}^{2} \left(\frac{k_p}{k},\frac{k_p}{k}\right)} \label{eq:anaLN}
\end{equation}
where $f(u,v)$ is the contraction of polarization tensor calculated as following:
\begin{equation}
       f(u,v) =  \frac{1}{16 \, u^{4}\,v^{4}} \left\{v^{8} + \left(u^{2} - 1\right)^{4} + 4 v^{6}\,\left(7 u^{2} - 1\right) + 4 v^{2} \left(u^{2} -1\right)^{2} \left(7 u^{2} - 1\right) + v^{4} \left(6-60 u^{2} + 70 u^{4}\right)\right\} \,.
\end{equation}
It is important to investigate the infrared tail of induced density power spectrum in \cref{eq:anaLN} as it provides the growth of the spectrum and plays an important role in formation of PBH formation. Infrared scales corresponds to $k \gg k_p$ and in such limit the spectrum reduces to the following:
\begin{equation}
    \mathcal{P}^{\rm LN}_{\delta^{(2)}}(k \ll k_p)  \overset{\Delta \rightarrow 0}{\approx} \frac{1}{\pi^3} \, \left(\frac{k}{k_p}\right)^{4} \, \left[6\,\left(\frac{k_p}{k}\right)^2 - 4 \, \log \left(\frac{k_p}{k} \right) - 3\right] \label{eq:IRLN}
\end{equation}
It is clear from \cref{eq:IRLN}, the induced power spectrum $\mathcal{P}^{\rm LN}_{\delta^{(2)}} \propto k^{4}$ with a peak situated around  $ k / k_p \approx \sqrt{3}$.

\begin{figure}[ht]
\includegraphics[scale=0.45]{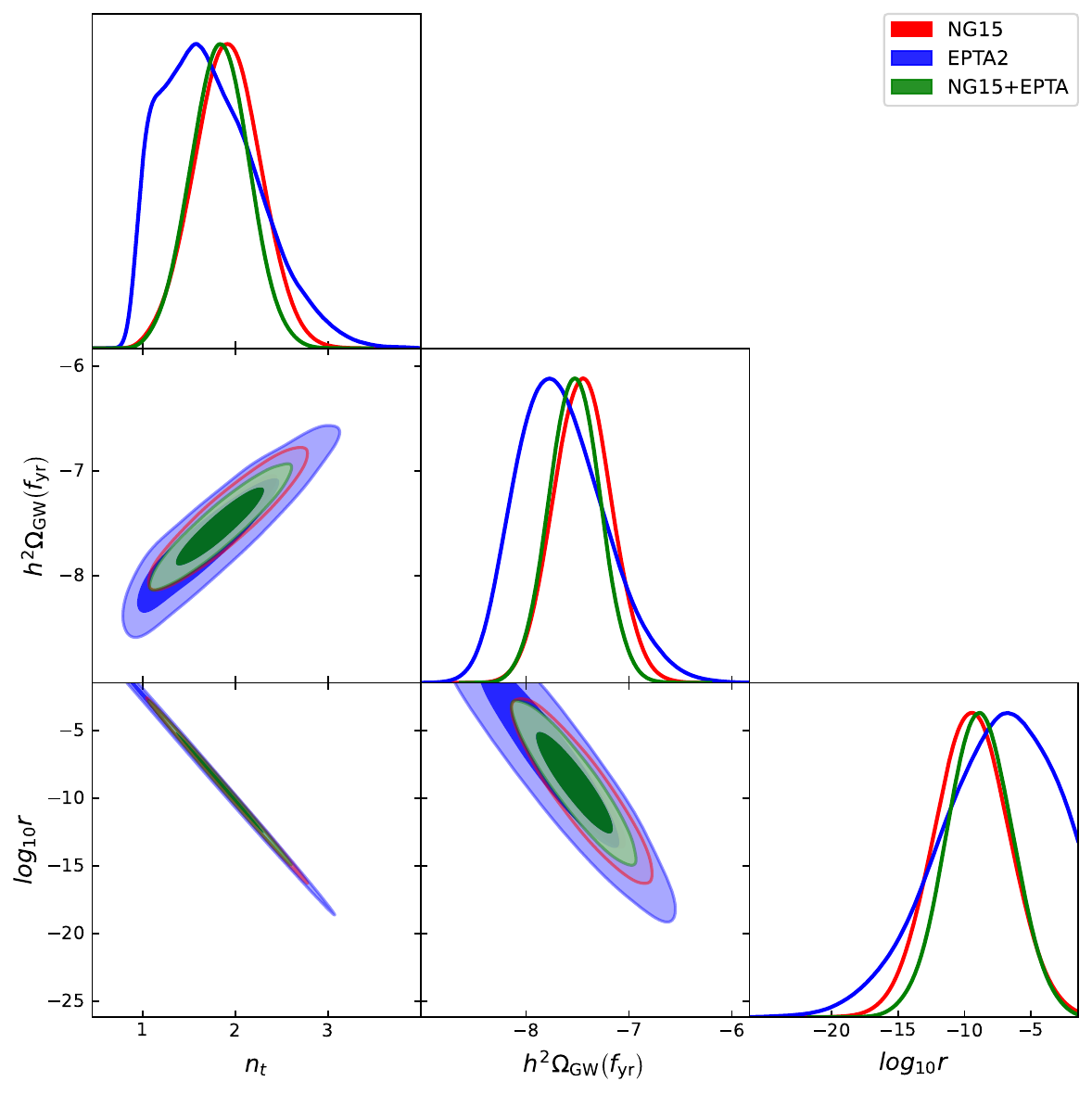}
\includegraphics[scale=0.45]{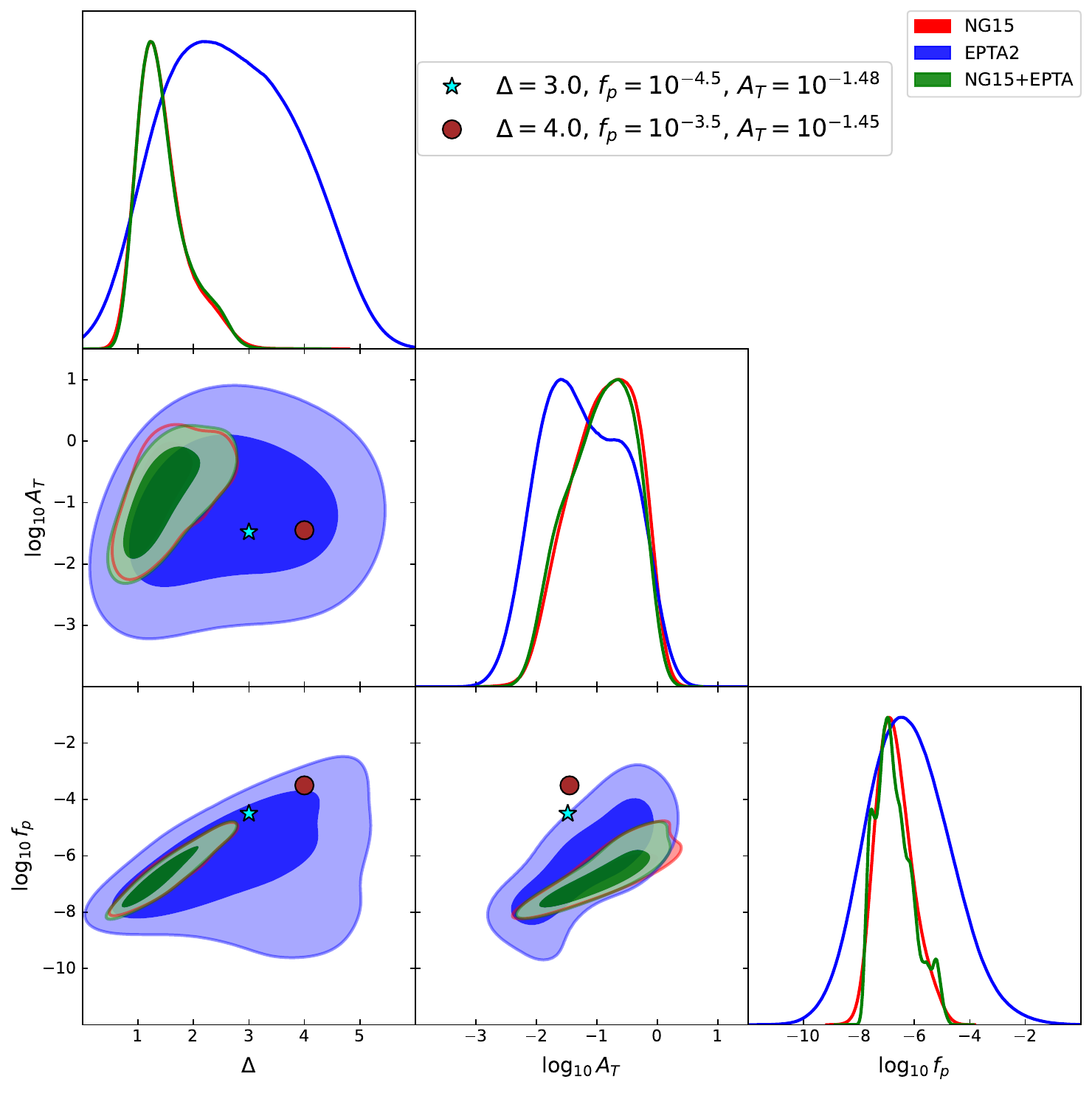}
\caption{Triangle plot for posterior distribution for PL (top) and LN (bottom) spectrum. Note that $\Omega_{\rm GW}(f_{\rm yr})$ is a derived quantity derived using the posterior values for $\log_{10} r$ and $n_t$. The brown and cyan points corresponds to PBHs with mass function dominated at $10^{-12} M_{\odot}$ and $10^{-10} M_{\odot}$  shown in main text.} \label{fig:triangle_PL}
\end{figure}

\section{Results for PGW energy spectrum from NANOGrav and PTA}. In the main text, we show the two dimensional contour plots for relevant model parameters. In this section, we detail the observational data and present the full posterior distributions. The NANOGrav data set measures the characteristic strain across 15 frequency bins spanning $f \in [2 \times 10^{-9}, 6 \times 10^{-8}]$ Hz, while the EPTA DR2 data provides measurements in 9 bins covering $f \in [3.0 \times 10^{-9}, 2.8 \times10^{-8}]$ Hz. We use \texttt{PTArcade} in ceffyl modes to generate the chains for both data individually. After generating the posterior chains, we perform the KDE to jointly analyze both measurements. We provide the triangle plots for model parameters obtained from the analysis of NANOGrav 15 and EPTA DR2 data. In the left panel of \cref{fig:triangle_PL} and \cref{fig:PLEN} (main text), we use a derived parameter defined as gravitational wave energy density evaluated at reference frequency $f_{\rm yr} (= 1 \rm yr^{-1} \approx 3.17 \times 10^{-8}$ Hz) given as 
\begin{equation}
    h^2\, \Omega_{\rm GW} \left(f\right) = 1.19 \times 10^{-16} \, \left(\frac{r}{0.032}\right) \,\left(\frac{A_s}{2.1 \times 10^{-9}}\right) \, \left(\frac{f}{f_s}\right)^{n_t}\,,
\end{equation}
with $f_{s} = 8 \times 10^{-17}$ Hz is frequency corresponding to the CMB pivot scale.

\section{Derivation of Log-Normal Power spectrum from a Inflationary model} \label{apn:LNspectrum}
We consider the primordial GWs from axion-gauge field dynamics \cite{Dimastrogiovanni:2016fuu} to illustrate the log-normal shape of the tensor power spectrum. The Lagrangian for Chromo-Natural Inflation (CNI) is given as:
\begin{equation}\label{cnii}
S_{\text{CNI}}=\int d^4x\sqrt{-g}\left[\frac{M_{\text{Pl}}^{2}}{2}R-\frac{1}{2}\left(\partial\chi\right)^{2}-U(\chi)-\frac{1}{4}F_{\mu\nu}^{a}F^{a\mu\nu}+\frac{\lambda\chi}{4f}F_{\mu\nu}^{a}\tilde{F}^{a\mu\nu}\right],
\end{equation}
where $M_{\text{Pl}}$ is the reduced Planck mass, $R$ is the Ricci scalar, $\chi$ is a pseudo-scalar axion field with potential $U(\chi)$, $F_{\mu\nu}^{a}\equiv\partial_{\mu}A^{a}_{\nu}-\partial_{\nu}A^{a}_{\mu}-g \epsilon^{abc} A^{b}_{\mu}A_{\nu}^{c}$ is the field strength tensor for an $SU(2)$ gauge field $A_{\mu}^{a}$, and $\tilde{F}^{a\mu\nu}\equiv\epsilon^{\mu\nu\rho\sigma}F^{a}_{\rho\sigma}/(2\sqrt{-g})$ is its dual.

Following the derivation in \cite{Thorne:2017jft}, the tensor power spectrum in the super-horizon limit is given by:
\begin{equation} \label{PQh}
\mathcal{P}_h^{(s)}(k) =\frac{\epsilon_B H^2}{\pi^2 M_{\rm Pl}^2}\mathcal{F}^2,
\end{equation}
with $\mathcal{F}^2 \equiv \left|\mathcal{F}_B+\sqrt{\epsilon_E/\epsilon_B}\mathcal{F}_E\right|^2$. Here, $H_{\inf}$ is the inflationary Hubble parameter, and $\epsilon_B \equiv g^2Q^4/(M_{\rm Pl}^2 H_{\inf}^2)$ quantifies the energy density in the SU(2) gauge field. The function $\mathcal{F}(m_Q)$ is monotonically increasing for $3\le m_Q \le 7$ and is empirically fitted by:
\begin{equation}
\mathcal{F}(m_Q) \simeq \exp\left[2.4308m_Q-0.0218m_Q^2-0.0064m^3_Q-0.86\right],
\quad \text{for } 3\le m_Q \le 7.
\label{F fitting}
\end{equation}
The dynamical parameter $m_Q(t)\equiv g Q(t)/H_{\inf}$ is evaluated at horizon crossing ($k\sim aH_{\inf}$).

Solving the background equations of motion for $\chi(t)$ and $Q(t)$ under the slow-roll approximation yields:
\begin{equation}
m_Q(t) =m_* \sin^{1/3}\left[\chi(t)/f \right],
\label{mQt}
\end{equation}
where $m_* \equiv \left(g^2 \mu^4/3\lambda H^4_{\inf}\right)^{1/3}$ represents the maximum value of $m_Q(t)$.

To determine the width of the GW spectrum, we expand $\chi(t)$ around its peak value at $t = t_{*}$ (where $\chi(t_{*})=\pi f/2$):
\begin{align}
\chi(t)&\simeq \frac{\pi}{2}f +\dot{\chi}_{*} (t-t_{*}) \nonumber \\
&\simeq f\left[ \frac{\pi}{2} + \frac{2\xi_{*}}{\lambda} H_{\inf}(t-t_{*})\right],
\label{chi expansion}
\end{align}
where $\dot{\chi}_{*} \equiv \dot{\chi}(t=t_{*})$, $\xi_{*}\equiv \lambda \dot{\chi}_{*}/(2fH_{\inf})$, and one finds $\xi_{*}\simeq m_{*}+m_*^{-1}$ in the slow-roll regime.
This expansion leads to an approximation for $m_{Q}(t)$ valid near its maximum:
\begin{equation}
m_Q(t)\simeq m_* \left[1-\frac{1}{6} \left( \frac{H_{\inf}(t-t_{*})}{\Delta N}\right)^2\right],
\qquad \text{(for } t\sim t_{*}),
\label{mQ approx}
\end{equation}
where we have defined $\Delta N \equiv \lambda/(2\xi_{*})$.

Substituting this into Eq.~\eqref{PQh} and using $H_{\inf}(t-t_{*})=\ln(k/k_p)$, we obtain the leading-order result:
\begin{equation}
\mathcal{P}_h^{(s)}(k) \simeq \frac{\epsilon_{B*} H_{\inf}^2}{\pi^2 M_{\rm Pl}^2}\mathcal{F}^2(m_{*}) \exp\left[-\mathcal{G}(m_{*})\frac{\ln^2(k/k_p)}{\Delta N^2} \right],
\end{equation}
with $\mathcal{G}(m_*)\approx 0.666+ 0.81m_*-0.0145m_*^2-0.0064m_*^3$.
Note that the contribution from $\epsilon_B(t)\propto m_{Q}^4(t)$ in the prefactor is included. We refer reader to \cref{eq:LN}for more details regarding the calculations Comparing this expression with the template in Eq. (12) of our paper, we identify the following one-to-one correspondence between the template parameters and the fundamental model parameters:
\begin{align}
    A_T &= \dfrac{\epsilon_{B*} H_{\inf}^2}{\pi^2 M_{\rm Pl}^2}\mathcal{F}^2(m_{*}) \\
    \Delta^{2} &= \Delta N^{2} / (2 \mathcal{G}(m_{*}))
\end{align}

\end{document}